\documentclass[12pt]{iopart}

\usepackage{iopams}  

\usepackage{latexsym}      
\usepackage{graphicx}      

\usepackage[compress,nospace]{cite}

\begin{document}

\title[Trends and statistics in name mentions in news]{Finding trends and statistical patterns in name mentions in news}

\author{AMC Jayin and RC Batac}

\address{National Institute of Physics, University of the Philippines Diliman, Quezon City 1101 Philippines}
\ead{rbatac@nip.upd.edu.ph}

\begin{abstract}
We extract the individual names of persons mentioned in news reports from a Philippine-based daily in the English language from 2010-2012. Names are extracted using a learning algorithm that filters adjacent capitalized words and runs it through a database of non-names grown through training. The number of mentions of individual names shows strong temporal fluctuations, indicative of the nature of ``hot'' trends and issues in society. Despite these strong variations, however, we observe stable rank-frequency distributions across different years in the form of power-laws with scaling exponents $\alpha = 0.7$, reminiscent of the Zipf's law observed in lexical (i.e. non-name) words. Additionally, we observe that the adjusted frequency for each rank, or the frequency divided by the number of unique names having the same rank, shows a distribution with dual scaling behavior, with the higher-ranked names preserving the $\alpha$ exponent and the lower-ranked ones showing a power-law exponent $\alpha' = 2.9$. We reproduced the results using a model wherein the names are taken from a Barabasi-Albert network representing the social structure of the system. These results suggest that names, which represent individuals in the society, are archived differently from regular words. 
\end{abstract}

\maketitle

\section{Introduction}

Many aspects of human societies have been viewed from a complex systems perspective, wherein humans are viewed as decision-making agents that interact with a finite number of contacts in a social network~\cite{BarabasiAlbertSCIENCE1999,NewmanPRE2001}. Previous works have quantified the emergent properties in economic~\cite{AmaralEtAlPHYSA2001}, political~\cite{FowlerSOCNET2006}, and social~\cite{BajardiEtAlEPJDS2015,BarabasiEtAlPHYSA2002} systems that result from the collective effect of such individual decisions of human agents, giving us a snapshot of the prevailing conditions in the society at a particular point in time. However, one inescapable aspect of human interactions is the fact that norms, trends, and other intangible features that define individual decisions change over time~\cite{LegaraEtAlIJMPC2009,WengEtAlSCIREP2012,WangHubermanEPJDS2012}. This temporal dependence, therefore, may need to be included in further studies of complex behavior in societies to better understand how it can change the dynamics of the system. To this end, we propose to study news reports, which can be viewed as daily records of the dynamics and interactions of the society. 

News reports, unlike many other personal publishing platforms like blogs and social media~\cite{FerraraEPJDS2012,AlisLimPLOSONE2013}, benefit from editorial review, which, in most cases, will ensure the veracity of the information recorded. In this work, we focus on print media, particularly news dailies, as they provide an additional constraint of having a finite space, thus ensuring that their contents are deemed to be the most important issues of the day. Previous works have mined sentiments~\cite{GodboleEtAl2007,LeetaruFM2011}, frames~\cite{LegaraEtAlEXPSYS2013} and various topic trends~\cite{MahajanEtAl2008,MontesEtAlCOMPUTACION2015} in news reports. These works, among others, have shown that news is an important source of data that can be extracted and analyzed to reveal patterns and trends in the complex interactions in human societies. 

However, as noted earlier, humans are the main agents that drive the complexity in the society. Why, then, do we find very little work on mining names in the news? We believe that two factors are mainly responsible for this. First, names are believed to be exhibiting the same characteristics as lexical or `non-name' words, so they are mined and analyzed together with them. This view, however, neglects the fact that lexical words have relatively longer lifetimes in the language. On the contrary, names represent actual individuals who not only have finite lifetimes but also have different chances to be mentioned in the news. Secondly, and perhaps even more challenging, naming conventions differ significantly across cultures, making a universal algorithm for extracting them quite difficult. 

Here, we extract the names of personalities as reported in an English-language daily from the Philippines. We choose this small subset for data extraction because of our familiarity with cultural naming conventions, making it easier for us to extract actual names and filter out other words that are of similar structures. We also propose guiding principles that may be useful for other authors in a similar search in a different cultural context. With our algorithm, we are able to extract around $10^5$ unique names over yearly periods for three years, 2010-2012, highlighted by a period of relative dynamism in the society due to the National Elections on May 2010. Our analyses reveal robust statistical properties despite the very dynamic trends in name mentions. Comparisons with similar statistical distributions of lexical words reveal subtle but important differences due to the nature of names as being representative of active agents and not just tools for communication. Finally, we present a simple model that is based on the assumed structure of society to explain the resulting statistical distributions. These results provide a quantitative description of societal dynamics and paint a picture of what society itself deems as important.

\section{Newspaper Data Mining and Name-Extraction Algorithm}

All news reports are collected from online archives of {\it The Philippine Star}~\cite{Philstar}, one of the major English-language dailies in the Philippines with a circulation of 275,000. The {\it Star} has six major sections: News, Features, Opinion, Business, Entertainment, and Sports. For our purpose, we use the general term ``news'' or ``articles'' to refer to the written material contained in the newspaper, regardless of its actual section. On average, for the years 2010 to 2012, there are 150 to 170 articles in every daily edition of the newspaper. 

We chose to extract the data sets from year 2010 up to 2012 for completeness and relative recency. Moreover, the year 2010 is highlighted by national elections for all elective positions in the Philippines. During this period, we expect to find a strong coupling of news reports with societal dynamics, such that the names of the persons deemed to be most important and influential (e.g. the presumptive candidates) are more often reflected in the news. 

As mentioned earlier, extracting names first require knowledge of the naming conventions used in a particular setting. In the Philippines, cultural conventions require a given name (which can be more than one word), a middle initial (from the first letter of the mother'€™s maiden name, or maiden name for married women), and a surname (usually from the father, or from the husband for married women) to completely identify an individual: {\tt Name M. Surname}. In some cases, the middle initial is dropped, especially for known personalities: {\tt Name Surname}. Still, in some cases, aliases are included in the names, especially for politicians: {\tt Name} ``{\tt Alias}'' {\tt Surname} or {\tt Alias Surname}. On top of these, some others adopt a suffix at the end of their names to continue using the same names across generations: ``{\tt Sr.}'', ``{\tt Jr.}'', ``{\tt III}'', ``{\tt IV}'', etc. These nuances in naming conventions had to be accounted for to come up with an algorithm that identifies the correct names of individuals. 

\begin{figure}[h!]
\centering
{\includegraphics[width=0.6\columnwidth]{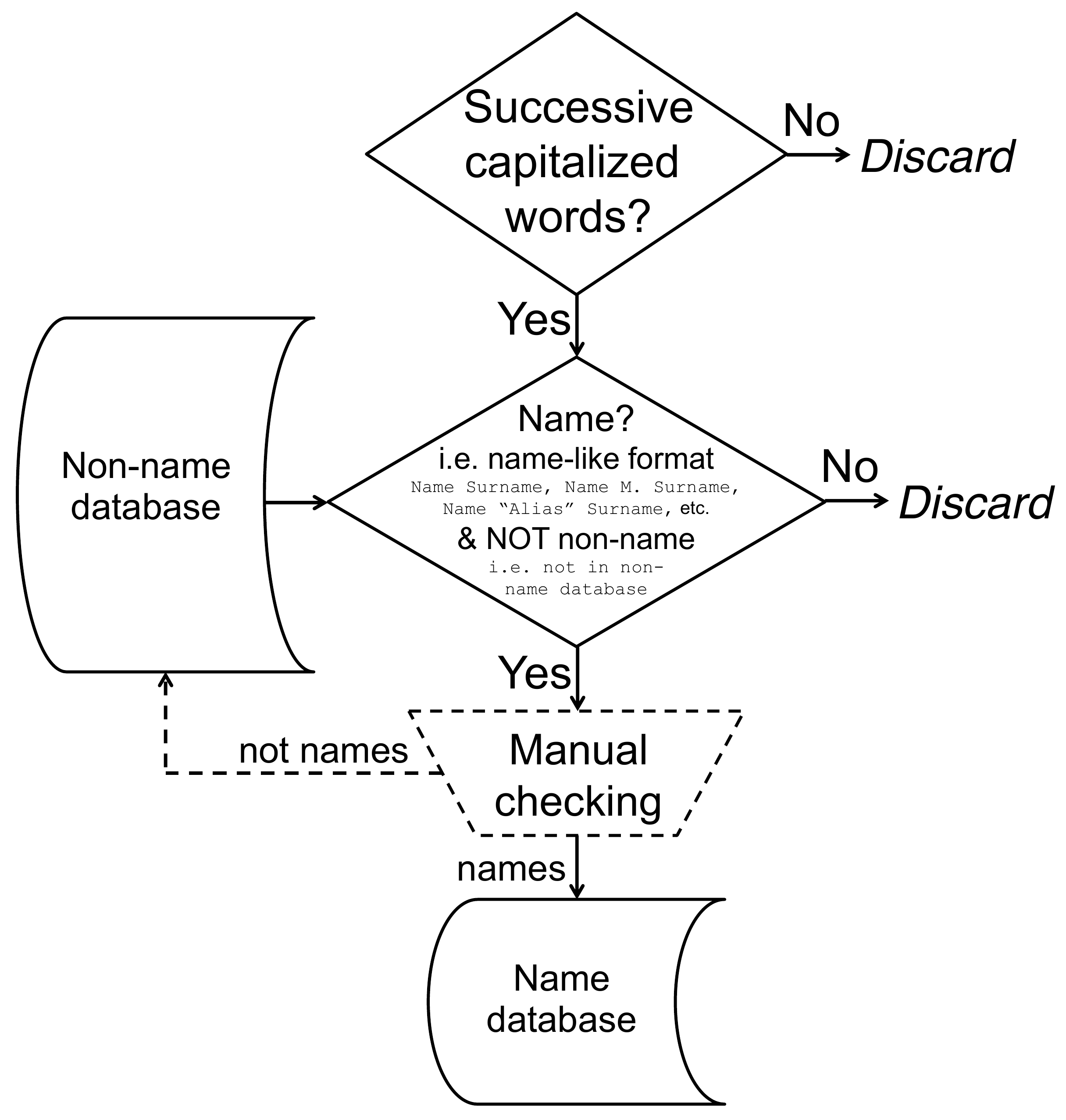}}
\caption{Algorithm for selection of names. Once successive capitalized words are extracted, another check is performed to verify if these are in name-like formats enumerated below. If yes, additional manual checking is done to confirm. Non-names, such as names of business firms, organizations, sports teams, or acronyms from government projects, are added into a non-name database that automatically filters out the instances in succeeding articles. Names are added into the name database and is used for data processing. The broken lines around `manual checking' indicate that this step is done decreasingly as extensive human intervention is only required during the first few months. Eventually, enough non-name words are obtained to cover most of these cases, allowing for an almost automatic classification.}
\label{Fig1algo}
\end{figure}

In Figure~\ref{Fig1algo}, we present a schematic representation of the algorithm used for such name identification. The algorithm is iterative, and requires continuous training and human intervention, albeit at a decreasing rate. The program first takes in the text file of the article and takes in successive words with capitalized first letters. Then, non-name capitalized words are removed, initially by human intervention. Examples of such non-name words that are manually removed are names of business firms, organizations, sports teams, or acronyms from government projects. These words, which are likely to come out again as longer periods are considered, are placed in a non-name database, which is eventually used by the algorithm to immediately filter out non-names in succeeding data. As more news files are processed, the list of non-name capitalized words also grows, resulting in the decrease in human intervention across longer periods of analyses. 

The names are then collected and grouped for different periods (in years) and newspaper sections (for 2010 data). Particular emphasis is given to the most mentioned individuals, as they are believed to be the main drivers of activity in the society over the period considered.

\section{Results}

\subsection{Robust rank-frequency distributions }

\begin{figure}[h!]
\centering
{\includegraphics[width=\columnwidth]{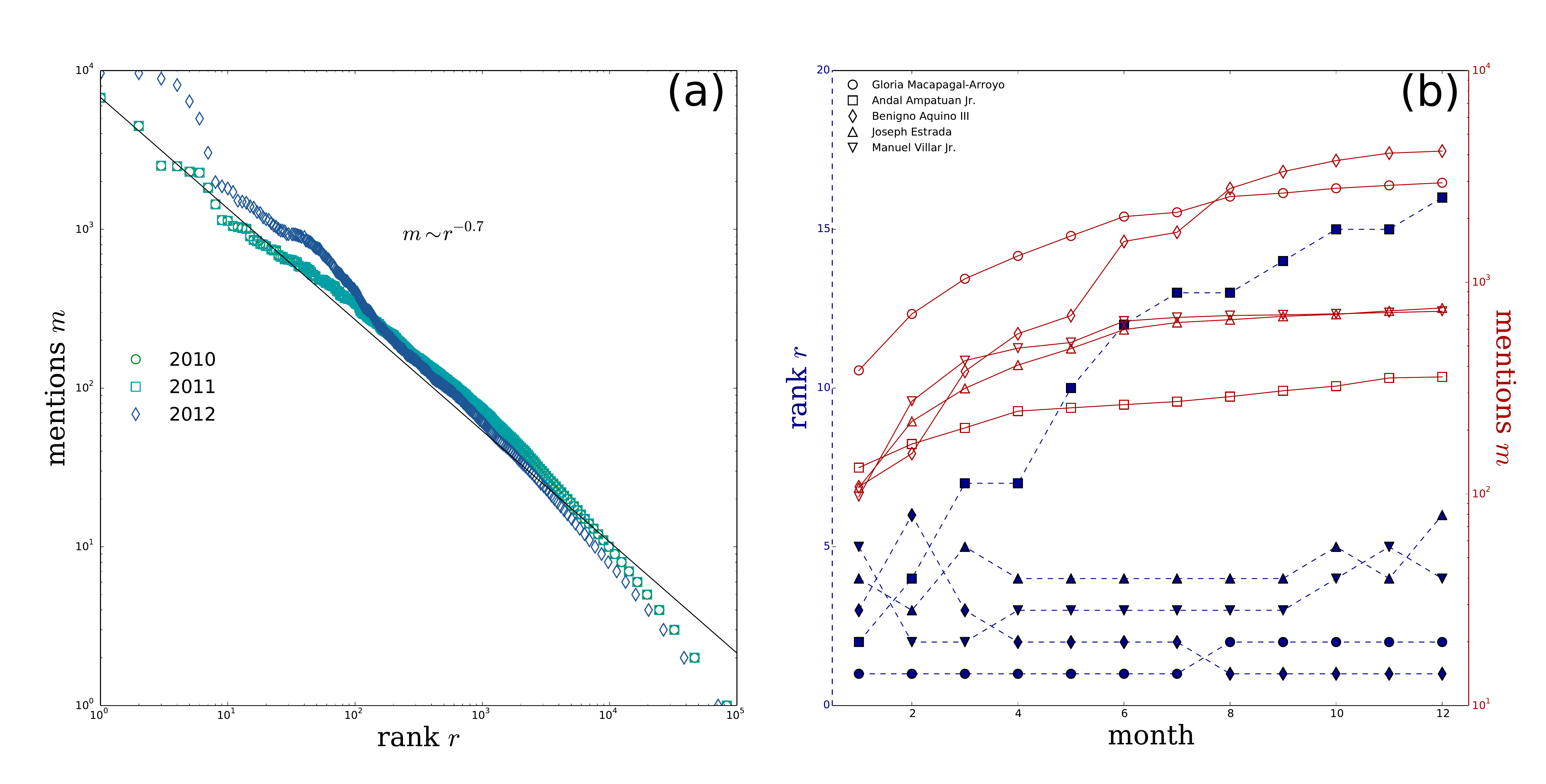}}
\caption{(a) Statistical distributions of the number of occurrence of a particular name, $m$, versus its rank $r$. The distributions, which are obtained for the different years 2010, 2011, and 2012, all show power-law behaviors $m \sim r^{-\alpha}$, where $\alpha = 0.7$. (b) Trends of representative top-ranked names. In January 2010, the following persons obtained the highest $m$ and are thus in the top 5 in $r$. Adding the corresponding occurrences across the different months result in changes in $r$ and $m$, showing the high variability of the ranking and actual mentions of names in news. }
\label{Fig2mr}
\end{figure}

Names, as in words, have different frequencies of occurrence in a text. Each name, therefore, may be characterized by its rank $r$, wherein $r = 1$ represents the most frequently occurring name. In lexical words, one of the most well known results is Zipf's law, which is the occurrence of power-law rank-frequency distributions $p(r) \sim r^{-\alpha_0}$, where $\alpha_0 = 1$~\cite{Zipf1949}. Zipf's law is found to be almost universal among many languages~\cite{CanchoSolePNAS2003, KanterKesslerPRL1995, RousseauZhangSCIENTO1992, DahuiaEtAl2005} and through diverse corpora gathered across different time periods~\cite{HaEtAl2002, Li1992}. To check whether Zipf's law is obeyed by names, which are also used in text but refers to actual individuals, we present in Figure~\ref{Fig2mr}(a) the number of occurrence of a particular name, $m$, as a function of its rank $r$ for names mined from the {\it Star} in 2010-2012. One readily observes that despite the similar power-law behavior that spans several decades, there are obvious differences in the statistical behavior with that of Zipf's law. 

The $m$ vs. $r$ distribution is shown to follow a power-law behavior $m \sim r^{-\alpha}$ where $\alpha = 0.7$. Consistent with our expectation that the {\it Star} will have a limited number of articles, the plots shown in Figure~\ref{Fig2mr}(a) almost collapse under the same curve even without normalization. The robust mentions vs. rank plots is reminiscent of Zipf's law, although the obtained scaling exponent $\alpha$ is not unity. This difference signifies that names, statistically speaking, may not be treated as lexical words as they appear in a text. This is in agreement with our view that names are not just tools of communication but are actually representing agents that affect changes in the society. However, the fact that the observed distributions are robust over different time periods is in itself remarkable, considering the temporal trends in name mentions in newspaper articles. 

Personal names, just like the people they represent, have finite lifetimes. Even for the most influential persons in history, the mentions in news articles decay after the death of the individual, but may crop up again intermittently during anniversaries and other special occasions. Moreover, even for current influential people, we find very irregular temporal trends in name mentions. Figure~\ref{Fig2mr}(b) shows the monthly tracking of the ranks $r$ [left broken axis] and cumulative mentions $m$ [right solid axis] for the top five most highly cited individuals in the {\it Star} for January 2010. Over the short period considered, both $r$ and $m$ show the temporal intermittency of ``€œpopularity''€ in news articles. Increases in $r$, and the corresponding decreases in steepness of $m$, denote a shift of focus of news reporting away from an individual. On the other hand, in some cases, some individuals gain news media mileage, as denoted by the decrease in the numerical value of $r$ and steeper increases of $m$ over the course of the month. In some cases, spikes denoting short-term popularity are observed for less popular individuals, resulting in wild fluctuations in the $r$ and $m$ trends. 

As a case in point, we note that the $r = 1$ individual for January 2010 is {\tt Gloria Macapagal-Arroyo}, then President of the Philippines. Also during this month, the $r = 3$ individual is {\tt Benigno Aquino III}, then a Senator but is one of the contenders for the Presidency. Before the elections, {\tt Aquino} experienced a fluctuation in ranks, even becoming $r = 6$ in March shortly before the elections and $r = 2$ after that. In May 2010, he won the Presidency and received heightened media coverage. After the Presidential Inauguration in June 30, 2010, the media shifted its attention to {\tt Aquino} as the new President, which led to his becoming the $r = 1$ name from August 2010 through the rest of the year. 

These observations about the temporal variations of the name mentions in news further highlight the significance of the power-law $m$ vs. $r$ distributions. Despite the very irregular $r$ and $m$ trends in time, the cumulative effect is to produce the same scale-free plots of the number of mentions vs. rank. This may be considered as an emergent feature that may be resulting from self-organization in the way the society reports individuals in news. Unlike lexical words, which are guided by the rules of grammar, names are reported based on the perceived relevance of the individuals they represent. The rank, therefore, is decoupled with the actual name; news will always feature the most influential individual more often, regardless of {\it who} that individual is.

\begin{figure}[h!]
\centering
{\includegraphics[width=\columnwidth]{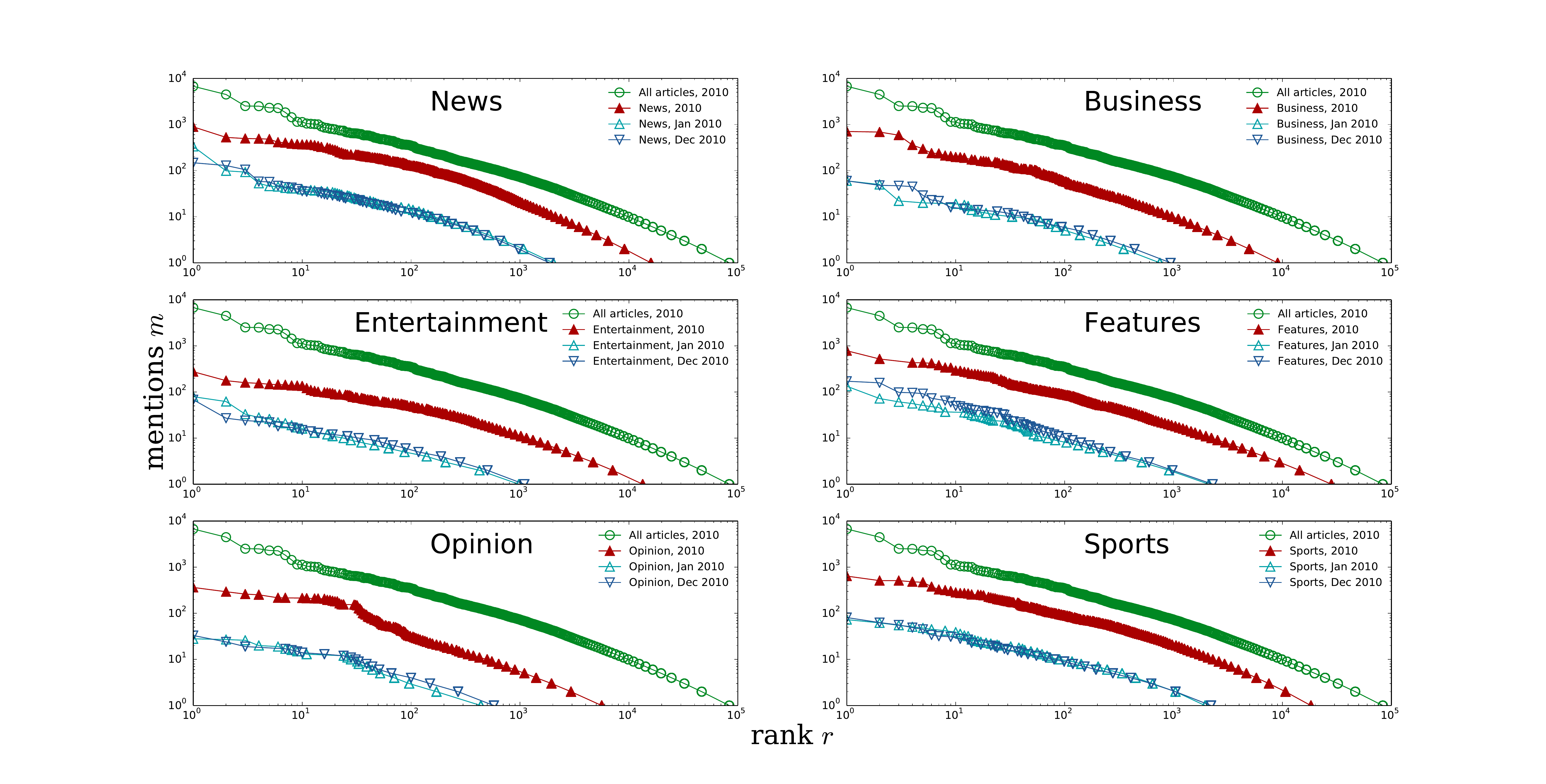}}
\caption{Statistical distributions of rank $r$ versus number of mentions $m$ for the six major sections of {\it The Philippine Star}. Each panel shows the name mentions for articles that appeared in: the entire paper in 2010 (\opencircle); that particular section in 2010 (\fulltriangle); and that particular section in January (\opentriangle) and December 2010 (\opentriangledown). The monthly rank-mention distributions show similar behavior for the same section, suggesting that the temporal robustness is indicative of the nature of each of the newspaper sections. However, the entire-year section distribution is generally not the same as the aggregate yearly distribution.}
\label{Fig3sec}
\end{figure}

Interestingly, this robustness in time is not observed for the different sections of the newspaper. As can be observed in Figure~\ref{Fig3sec}, when the data is divided into the different sections, the resulting distributions have similar temporal trends but do not always show a similar behavior as the aggregate dataset of names. This can be attributed to the fact that the different sections have different sizes; in lexical words, the size of the corpus is an important factor that determines the resulting statistical distribution. On the other hand, we also believe that this is indicative of the nature of the individual newspaper sections. For example, the Opinions section may tend to focus on the important issues as perceived by the columnists, and thus involves relatively fewer individuals; this results in less steep distributions (i.e. comparable number of mentions) especially for the highest-ranked individuals. These per-section differences, however, are lost when the entire newspaper statistics is collected. This further highlights the robustness of the obtained power-law rank-frequency distributions. 

\subsection{Highest-rank uniqueness  and dual scaling}

\begin{figure}[h!]
\centering
{\includegraphics[width=\columnwidth]{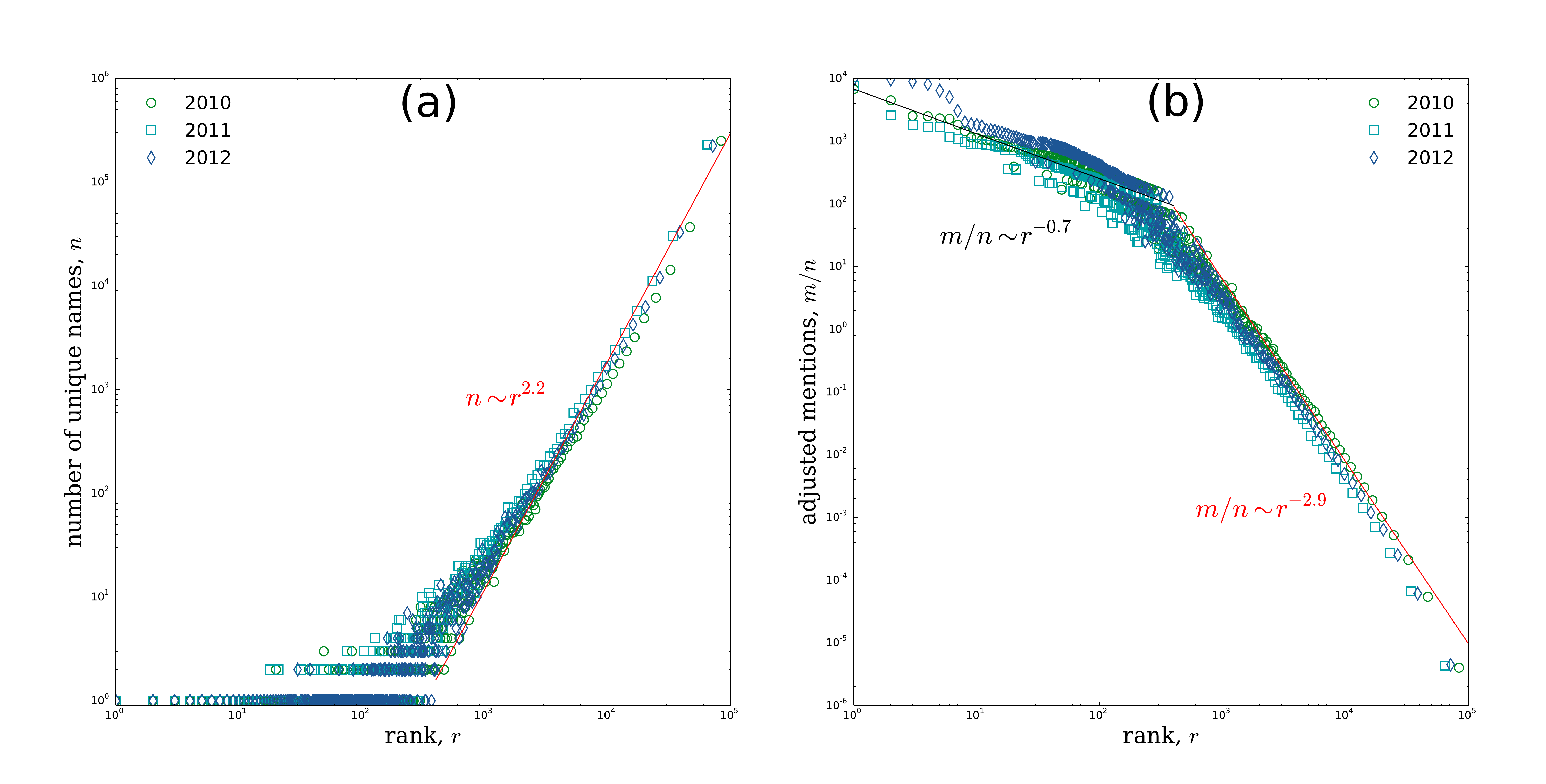}}
\caption{(a) Number of unique names $n$ per rank $r$. The plots show two regimes: for the first few ranks, the number of names are almost unique; after $r^\ast \approx 368$, the plot follows a power-law trend $n \sim r^{\beta}$, where $\beta \approx 2.2$. (b) Adjusted mentions $m/n$ vs. rank $r$. Due to the two regimes in $n$, the plot shows dual-scaling behavior: for $1 \leq r \leq r^\ast$, the scaling exponent is still $\alpha$, but for $r > r^\ast$, the new scaling exponent is $\alpha' = \beta + \alpha \approx 2.9$.}
\label{Fig4uniqueness}
\end{figure}

In Figure~\ref{Fig2mr}(a), we observe that there is a high level of asymmetry in the reporting of individuals in news, as quantified by the number of mentions of a name of certain rank. Another metric that clearly shows the disparity in the name reporting is on the number of unique names $n$ that correspond to a particular rank $r$. This is easily appreciated when considering the lowest ranks (i.e. highest $r$ values): many names are mentioned only once for a particular period due to the incidental nature of their activities e.g. crime victims and perpetrators, contest winners, and even obituary entries. In Figure~\ref{Fig4uniqueness}(a), we observe two regimes in the $n$ vs. $r$ plot that distinguishes the individuals based on their relative popularity by news mentions. 

Interestingly, the highest-ranked (lowest values of $r$) names show a regime of uniqueness: there is only one name that corresponds to that particular rank. In Figure~\ref{Fig4uniqueness}(a), this is shown by the almost flat region that extends from $r = 1$ to some $r^\ast$, where $r^\ast$ is the highest $r$ value with $n = 1$. For the different years considered, we find that this highest-valued unique rank is almost the same, $r^\ast = \{297, 289, 368\}$ for the years 2010-2012, respectively. Beyond this $r^\ast$, the $n$ vs. $r$ plot begins to show power-law behavior, which is comparable to the trend $n(r) \sim r^\beta$, where $\beta = 2.2$. We surmise that these two regimes emerge from the preferential nature of news reporting, and the finite number of articles of the newspaper. Because of the limited space, editors will tend to focus on the most important newsmakers, creating the uniqueness regime for low $r$ names. The number of names with limited mentions, on the other hand, tend to accumulate daily, resulting in the power-law regime for high $r$ values.

This division, in turn, can be incorporated in the statistical distributions of each ranks. Figure~\ref{Fig4uniqueness}(b) shows the plot of the adjusted mentions $m/n$ vs. the rank $r$, which shows two scaling regimes that are separated by $r^\ast$. The low values of $r$ preserve the scaling exponent $\alpha \approx 0.7$ because of their uniqueness. On the other hand, beyond $r^\ast$, we observe a scaling $m/n \sim r^{-\alpha'}$, with $\alpha' = \beta + \alpha \approx 2.9$. The faster decay for $r > r^\ast$ denotes the higher level of inequity among the individuals with fewer mentions, suggesting that the adjusted mentions $m/n$ may be used as a better measure of their significance in the news reports. 

Interestingly, similar dual-scaling regimes are observed in lexical words~\cite{CanchoSoleJQLIN2001,MontemurroPHYSA2001,PetersenEtAlSCIREP2012,GerlachAltmannPRX2013}. Gerlach and Altmann explained this result by using a stochastic model that identifies words as being ``core'' or ``non-core'' based on their usage~\cite{GerlachAltmannPRX2013}. Here, we report a comparable dichotomy which can be gleaned from Figure~\ref{Fig4uniqueness}(a) and (b): the regime $1 \leq r \leq r^\ast$ is the regime of the ``popular'' newsmakers (similar to the ``core'' words in lexical analyses) while those with $r > r^\ast$ are the ``not popular'' ones (similar to the ``non-core'' words). As noted earlier, however, despite these similarities, the mechanism of news selection is fundamentally dissimilar to that of word usage. Whereas grammatical structure dictates the occurrence of words, and, especially for core words, temporal variations occur over very long periods, names are highly dynamical, and the most frequently occurring name may be different for different time periods. 

\section{Model of name statistics: Barabasi-Albert network sampling}

Many previous works have shown that the social network structure underlying the society can be modelled using a Barabasi-Albert (BA) network~\cite{BarabasiAlbertSCIENCE1999}, wherein the number of links $k$ obeys a power-law distribution $p(k) \approx k^{-\delta}$, where $\delta \approx 3$. The BA network is deemed to be close to the mechanisms of social connections in the real world, which are governed by preferential attachment mechanisms. 

News may be viewed as a reflection of the societal structure at any given point in time. Therefore, we may think of the process of news reporting as a mechanism of fetching names of individuals and their social connections from a social network structure. Guided by this simple argument, we propose a model for generating the statistics obtained from data, by using a BA network where the nodes represent individuals and links represent their connections in the society. From this BA framework, ``news reporting'' is conducted by sampling the network, i.e. taking a node and all of its connections and giving them a unit mention. As a first approximation, this relies on the fact that the most closely-linked individuals in the society are more likely to be mentioned together in a news article. This sampling is done for $T$ iterations, which, for news, will correspond to the total number of articles per day multiplied by the period (in number of days) of accumulated data. 

\begin{figure}[h!]
\centering
{\includegraphics[width=\columnwidth]{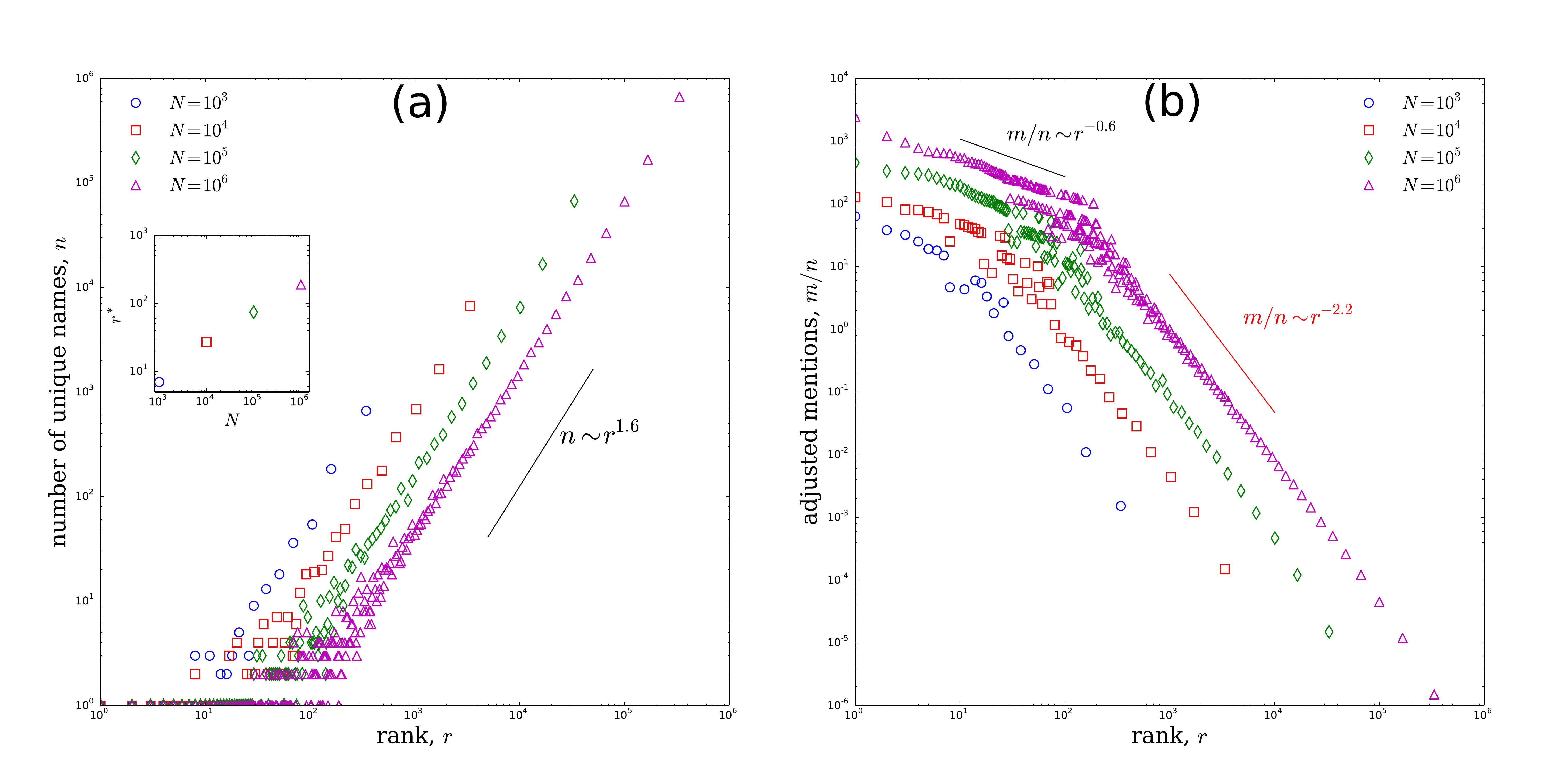}}
\caption{(a) Number of unique names $n$ per rank $r$ for different BA network sizes $N = \{10^3, 10^4, 10^5, 10^6\}$. The inset shows $r^{\ast}$ for the different BA network sizes. We observe that as the network size increases, $r^{\ast}$ shifts to higher ranks, implying that the number of unique names also increases. However, the power-law exponent $\beta_{BA} = 1.6$ is consistent regardless of network size, showing that it is ultimately linked with the BA network structure. (b) Distribution of rank $r$ versus the adjusted number of mentions $m/n$ for different network sizes, with $\alpha_{BA} = 0.6$ for $1 \leq r \leq r^{\ast}$ and $\alpha'_{BA} = 2.2$ for $r > r^{\ast}$. The relationship $\alpha'_{BA} = \beta_{BA} + \alpha_{BA}$ holds regardless of network size.}
\label{Fig5BA}
\end{figure}

In Figure~\ref{Fig5BA}, we present the statistics obtained from the simplest implementation of the said model, for the case of a BA network with $N$ nodes, and with each node sampled once, corresponding to $T = N$. In this case, the total number of mentions of each of the nodes is $m = 1 + k$, which corresponds to the number of times a node and its connections are sampled. Interestingly, we recover similar statistical properties as observed from the data. 

The regimes of uniqueness for low $r$ values and the corresponding power-law trend for higher $r$ values are shown in Figure~\ref{Fig5BA}(a). The power-law regimes all follow similar trends comparable to $n(r) \sim r^{\beta_{BA}}$, where $\beta_{BA} \approx 1.6$. The increasing trend of $r^\ast$ vs. $N$, as shown in the inset of Figure~\ref{Fig5BA}(a), indicates that larger system scales (i.e. larger BA network and/or longer sampling time) lengthens the uniqueness regime, making the system more disparate; the effective mentions for the more ``popular'' individuals increase with system scale. In Figure~\ref{Fig5BA}(b), we observe the dual-scaling regime similar to the ones observed in data, characterized by the exponents $\alpha_{BA} \approx 0.6$ and $\alpha_{BA}' = \beta_{BA}+\alpha_{BA} \approx  2.2$.

The apparent universality of the power-law exponents $\alpha_{BA}$, $\beta_{BA}$, and $\alpha'_{BA}$ with system scale indicates that these exponents result from the structural properties of the BA network, which is the one thing kept constant throughout the simulations. Because of the similarities in the trends and in the actual values of the exponents obtained with the data, we believe that the results obtained from news mentions in Figure~\ref{Fig4uniqueness}(a) and (b) are also derived from the underlying societal structure, which is just reflected in the finite space of the print news. 

\section{Conclusion}

Guided by our assumption that news is primarily about individuals and their actions, we have mined individuals' names as mentioned in the news using a learning algorithm that incorporates the cultural intricacies in naming conventions. We have shown that, similar to lexical (non-name) words, names follow robust power-law rank-frequency distributions akin to Zipf's law; on the other hand, the difference in exponents are deemed to be a reflection of the fact that names are not just language tools but are agents that affect change in the society. The robustness of the distributions is particularly noteworthy, given the highly dynamic nature of relative popularity of individual names in news. 

Another interesting result is the high disparity in the number of names corresponding to a particular rank. Higher-ranked individuals are found to be almost unique, while those with lower ranks have power-law increasing numbers per rank. This dichotomy results in the dual-scaling regime for the effective mentions of individuals, which was also observed for lexical words. We have presented a model for replicating the statistical distributions obtained, showing that such distributions are fundamentally attributable to the network structure of society. 

We believe that our analyses and results form a crucial bridge between studies that focus on complex network structures of society and those that mine data from written records of such human systems. The study has also filled a relative research void by its use of news as corpora and individuals' names as data sets. The latter point on probing name mentions may be further investigated using similar approaches that utilize other corpora, particularly for social media and other personal publishing platforms, to be able to account for similarities and differences.

\end{document}